\documentstyle[12pt]{article}

\topmargin -\headheight

\begin{document}

\title{The Importance of being Odd}
\author{ Yu.~Stroganov \\\\
{ Institute for High Energy Physics}\\
{ Protvino, Moscow region, Russia}}
\date{}

\maketitle

\begin{abstract}
In this letter I mainly consider a finite XXZ spin chain with
periodic boundary conditions and an \bf{odd} \rm number of sites. 
This system is described by Hamiltonian 
$   H_{xxz}=-\sum_{j=1}^{N}\{\sigma_j^{x}\sigma_{j+1}^{x}
   +\sigma_j^{y}\sigma_{j+1}^{y}
   +\Delta \>\sigma_j^z\sigma_{j+1}^z\} $.
As it turns out, the ground state energy is proportional
to the number of sites $E=-3N/2$ for a special value of the 
asymmetry parameter $\Delta=-1/2$. 
The trigonometric polynomial $q(u)$, the zeroes of which are
parameters of the ground state Bethe eigenvector, is
explicitly constructed. This polynomial of degree $n=(N-1)/2$
satisfies the Baxter T-Q equation. Using the second independent
solution of this equation that corresponds to the same eigenvalue
of the transfer matrix, it is possible to  find a derivative
of the ground state energy w.r.t. the asymmetry parameter.
This derivative is closely connected with the correlation
function $ <\sigma_j^z\sigma_{j+1}^z> =-1/2+3/2N^2 $. This 
correlation function is related to the average number of spin
strings for the ground state
$ <N_{string}> = \frac{3}{8}(N-1/N)$. I would like to stress
that all the above simple formulas are \bf not applicable to \rm 
the case of an even number of sites which is usually
considered.
\end{abstract}

\hfill I did not care what it was all about. 

\hfill All I wanted to know was how to live in it.

\hfill (Ernest Hemingway)

About 30 years ago Baxter noticed~\cite{BXYZ} that in some 
exceptional cases the ground state energy of the XYZ spin
chain, which has Hamiltonian
\begin{equation}
   \label{H2}
   H_{xyz}=-\sum_{j=1}^{N}
  \{J_x \sigma_j^{x}\sigma_{j+1}^{x}
   +J_y \sigma_j^{y}\sigma_{j+1}^{y}
   +J_z \sigma_j^z\sigma_{j+1}^z\}, \qquad 
   \vec\sigma_{N+1}=\vec\sigma_{1},
\end{equation}
has the especially simple value 
$$\lim_{N \rightarrow \infty} E/N=-(J_x+J_y+J_z),\  \mbox{
if}\ J_x,J_y\ \mbox{and}\ J_z\  \mbox{satisfy} 
\ J_x J_y + J_y J_z + J_z J_x =0.$$
 in the thermodynamic limit. He later noted~\cite{Baxter}, that
 the inversion relation gives a very simple eigenvalue for
 the 8-vertex model transfer matrix that  corresponds to (\ref{H2}).
Using standard notation for the Boltzmann weights of
the 8-vertex model we can reformulate the Baxter's
remark as follows.
If the weights satisfy the condition
\begin{equation}
   \label{C}
  (a^2+a b) (b^2+a b) =(c^2+a b) (d^2+a b),
\end{equation}
then the transfer matrix has (up to a sign) eigenvalue
$T =  (a+b)^N$. Consequently Hamiltonian (\ref{H2}), even for
 finite chains, has an exact eigenenergy of $E=-N(J_x+J_y+J_z)$.
A natural suggestion follows. Suppose that the corresponding
eigenvector is the ground state vector, then it is probably
possible to obtain interesting information for the finite chains. 
However, there are some problems. It is evident that the Baxter's
remark is valid for N=1 when both eigenvalues of the transfer
matrix are $T=a+b$, but for $N=2$, the transfer matrix does not
have the eigenvalue $T = (a+b)^2$.
For simplicity let us consider the 6-vertex model which is
the trigonometric limit ($d=0$) of the 8-vertex model.
Hamiltonian (\ref{H2}) is reduced to the XXZ Hamiltonian
with a special asymmetry parameter $\Delta=-1/2$:
\begin{equation}
   \label{H1}
   H_{xxz}=-\sum_{j=1}^{N}
  \{\sigma_j^{x}\sigma_{j+1}^{x}
   +\sigma_j^{y}\sigma_{j+1}^{y}
   -\frac{1}{2}\sigma_j^z\sigma_{j+1}^z\}, \qquad 
   \vec\sigma_{N+1}=\vec\sigma_{1}.
\end{equation}
Solving Baxter's T-Q equation for $N=2$ we easily
find a trigonometric polynomial $Q(u)$ of degree 2, but the
corresponding Bethe vector does not exist.

 There are at least \bf three ways \rm to fix 
 this problem and to obtain a simple eigenvalue
 in the transfer matrix spectrum. 
 The first two have to do with modifying system (\ref{H1}).
 
\bf Firstly \rm one can modify the boundary conditions.
In 1987 Alcaraz  et al.~\cite{ABBBQ}
considered an open XXZ spin chain with
a special magnetic field at the boundaries.
The authors carried out an intensive investigation
of the system. For $\Delta = -1/2$ they found
a linear dependence between the ground state energy 
and the number of sites.
The Hamiltonian of this chain can be expressed with the help
of elements of the Temperlay-Lieb algebra.
For $\Delta = -1/2$ this algebra has the trivial
one-dimensional representation.
In the circumstances it is possible to explain the simplicity
of the ground state energy. 
At present, due to multiple investigations in nineties,
(see, for example, the paper of Hinrichsen et al.~\cite{HMRS}
and the paper of Martin-Delgado and Sierra~\cite{MG}) this system
is considered trivial.
I intend to discuss all these questions in a future publication.
Now I limit myself to a reference to paper~\cite{FSZ},
where an exact solution of the Baxter's T-Q  equation 
was found. It is of importance that this solution corresponds
to the ground state of the system.

\bf Secondly \rm one can apply an additional horizontal field.
This field breaks the spin reversal  
invariance of the original model. As a result, a lot of new 
Bethe states emerge. For a special value of the field
strength one finds in the spectrum of the transfer matrix the 
simple value $T=-(a+b)^N$ (for even N only).
The associated spin chain Hamiltonian is described by Perk
and Schultz~\cite{PS}. I am indebted to Bazhanov and Baxter
who informed me about this possibility. 
It was investigated in the sixth section of paper 
~\cite{FSZ2}, which is an extended version 
of paper~\cite{FSZ}. 
 
The two above mentioned methods deal with the trigonometric case
only. I chose a third way.
\bf As it turns out, it is enough to consider
the usual $XYZ$ ($XXZ$) - spin chain with periodic boundary
conditions, but with an ODD  number of sites $N=2n+1$.
\rm I have checked that for $N=3,5,7$ the transfer
matrix for 8-vertex model has the largest eigenvalue
$T=(a+b)^N$ when the weights satisfy condition (\ref{C}). 

Due to technical difficulties typical for
8-vertex model, I limit myself here with
the trigonometric case ($d=0$).
The existence of the above-mentioned simple eigenvalue
for the ground state energy was discovered in the trigonometric
case by Alcaraz, Barber and Batchelor~\cite{ABB}.

In this case the 8-vertex model reduces to the 6-vertex 
model and formula (\ref{C}) reduces to 
$c^2=a^2 + a b + b^2$.
The corresponding Hamiltonian is given by (\ref{H1}).
\bf The simplicity of the ground state in this case allows
one to find simple explicit formulas for some of the correlations.
\rm

Let us consider a product of z-axis spin projection operators
acting at the neighbouring sites.
One of the simplest correlations is the average of this product
over the ground state 
\begin{equation}
\label{cpar}
C^{\parallel}_1(\Delta) \equiv  \> <\sigma^z_j \sigma^z_{j+1}>.
\end{equation}
Due to translation invariance, this correlation does not depend
on $j$. Similar correlations related to the x and y 
axes do not depend on $j$ also.
Their values coincide, due to z-axis rotation invariance.
Below we use the notation 
$$
C^{\perp}_1(\Delta) \equiv \> <\sigma^x_j \sigma^x_{j+1}>\>=
\><\sigma^y_j \sigma^y_{j+1}>.
$$

\bf The  simple formulae for these correlations,
valid for $\Delta=-1/2$ and for odd N, are 
the main result of the paper. 
\rm

Firstly let us formulate the starting point of our calculations.
It is well known that the Boltzmann weights of
the 6-vertex model can be conveniently 
parametrized by spectral parameter u and 
crossing-parameter $\eta$ as 
\begin{equation}
 \label{weights}
 a=\sin (u + \eta/2),\>\> b=\sin (u - \eta/2)\>\> \mbox{
 and}\>\> c=\sin \eta.
\end{equation}
The asymmetry parameter is related to
the crossing parameter via $\Delta=\cos \eta$.
For the special case $u=\eta/2$, transfer matrix $\hat T$
is proportional to a shift in the chain by one site  and its
eigenvalues for all states with zero momentum are $\sin ^N \eta$.
It is also known that the Hamiltonian for $XXZ$ spin chain
is related to the logarithmic derivative of the transfer 
matrix w.r.t. the spectral parameter:
\begin{equation}
   \label{HT}
  H_{xxz}=-\sum_{j=1}^{N}\{\sigma_j^{x}\sigma_{j+1}^{x}
  +\sigma_j^{y}\sigma_{j+1}^{y}
  +\cos \eta \>\sigma_j^z\sigma_{j+1}^z\} = N \cos \eta - 
  2 \sin \eta \biggl(\frac
  {\hat T^{\prime}_u}{\hat T} \biggr)_{u=\eta/2}.
\end{equation}
We will also use Baxter's T-Q equation 
\begin{equation}
    \label{TQ}
  T(u)\> Q(u) = \sin ^N (u + \eta/2)\> Q(u - \eta) +  
  \sin ^N (u - \eta/2)\> Q(u + \eta),
  \end{equation}
where $T(u)$ and $Q(u)$ are eigenvalues of transfer matrix
$\hat T$ and of auxiliary matrix $\hat Q$,
corresponding to a common eigenvector.
In the trigonometric case
\begin{equation}
\label{Qprod}
Q(u) = \prod^m_{j=1} \sin (u - u_j).
\end{equation}
where $u_j$ satisfy the Bethe equation.

\bf Firstly \rm, 
we explicitly find for an odd value of N
and for a fixed value of the crossing 
 parameter $\eta=2\pi/3$ two solutions
$Q(u)$ and $P(u)$ of equation (\ref{TQ})
corresponding to the  hypothetical
 eigenvalue of the transfer matrix
 $T(u)=(a+b)^N$.
We argue that the Bethe vector,
constructed with the help of $Q(u)$, corresponds 
to the ground state.
\bf Then \rm, using the result of paper~\cite{FSZ2},
we formulate the relation between
these two solutions and the derivative of 
the largest eigenvalue of the transfer
matrix $T(u)$ w.r.t. $\eta$.
\bf Lastly \rm, the knowledge of this 
derivative lets us find the correlation (\ref{cpar}).
 
Let us momentarily fix the crossing parameter 
$\eta=2\pi/3\  (\Delta=-1/2)$
 and consider the conjectured value $T=(a+b)^N$.
 Using parametrization (\ref{weights}), one can write this 
 as $T(u)=\sin ^N u$. Baxter's equation (\ref{TQ})
 for odd N takes the cyclic form 
 \begin{equation}
  \label{eqf}
  f(u)+f(u+\frac{2\pi}{3})+f(u+\frac{4\pi}{3}) = 0,
 \end{equation}
 where $f(u) = \sin^{2n+1} u \>\> Q(u)$.
To  solve the equation we follow \cite{FSZ}.
 Using cross-invariance of the T-Q equation and the 
 simple structure of the transfer matrix for $u=\eta/2$,
 one can show that $Q(u)$ is an even function and thus
 $f(u)$ is an odd trigonometric polynomial of degree $3n+1$,
 satisfying periodicity $f(u+\pi) = (-1)^{n+1}\>f(u)$.
 We can therefore write
 $$
 f(u) = a_1 \sin (3n+1) u + a_2 \sin (3n-1) u + a_3 \sin (3n-3) u \dots
 $$
 Equation (\ref{eqf}) is satisfied if $a_{3\nu}=0$.
 This condition implies
 $$
 f(u) = \sum_{k=0}^{n} \alpha_k \sin (1-3n+6k)u.
 $$
 The polynomial $f(u)$ is divided by $\sin ^{2n+1} u$
 by definition.
 This fixes the coefficients $\alpha_k$.
 It is clear that the first $2n$ derivatives have to be zeroes
 for $u=0$.
 The derivatives of even order are zeroes trivially while 
 the odd derivatives give
 $$
  \sum_{k=0}^{n} \alpha_k (1-3n+6k)^{2\mu +1}=0,\qquad \mu=0,1,\dots,n-1.
 $$
 This system is equivalent to the condition that the relation
 \begin{equation}
 \label{zero}
  \sum_{k=0}^{n} \alpha_k (1-3n+6k) P((1-3n+6k)^2)=0,
 \end{equation}
 is valid for all polynomials $P(x)$ of degree $n-1$.
 Let us consider $n$ polynomials
 of degree $n-1$:
 $$
 P_r(x) = \prod_{k=1,k \ne r}^{n} (x-(1-3n+6k)^2).
 $$
 Using these polynomials in formula (\ref{zero}) 
 we get the relations connecting the $\alpha_r$ with 
 $\alpha_0$.
 It is possible to write the answer in terms of binomial 
 coefficients
 \begin{equation}
 \label{f}
 f(u) = f_0 \sum_{k=0}^{n} 
 \biggl(\begin{array}{c}
 n - \frac{1}{3} \\
 k \end{array}\biggr)
 \biggl(\begin{array}{c}
 n + \frac{1}{3} \\
 n - k
 \end{array}\biggr)\sin(1-3n+6k)u,
 \end{equation}
 where $f_0$ is an arbitrary constant.
 
 The auxiliary function $g(u) = \sin ^{2n+1} P(u)$
 corresponding to the second independent solution
 of T-Q equation can be found by analogy
 \begin{equation}
 \label{g}
 g(u) = g_0 \sum_{k=0}^{n} 
 \biggl(\begin{array}{c}
 n - \frac{2}{3} \\
 k \end{array}\biggr)
 \biggl(\begin{array}{c}
 n + \frac{2}{3} \\
 n - k
 \end{array}\biggr)\sin(2-3n+6k)u.
 \end{equation}
 One can easily check that $f(u)$ and
 $g(u)$ satisfy ODE:
 \begin{eqnarray}
 \label{ODE}
 &f^{\prime \prime} - 6n \cot 3u \> f^{\prime} + (1-9n^2)f=0, \nonumber \\
 &g^{\prime \prime} - 6n \cot 3u \> g^{\prime} + (4-9n^2)g=0. 
 \end{eqnarray}
 It happens that occasionally these equation are more convenient
 in  calculations than explicit formulae (\ref{f}) and (\ref{g}).

The integer $m$ in (\ref{Qprod}) is equal to the number of 
reversed spins and related to the z-axis projection of the
total spin $S_z = N/2-m$.
 The solution $Q(u)$ has degree $m=n=(N-1)/2$ consequently 
the corresponding eigenvector has $S_z=1/2$.
In principle we can  construct this vector using 
 QISM \cite{QISM,QISM2}.  
It is also known  that  for 
the antiferromagnetic XXZ chain with an even N the ground state
has $S_z=0$~\cite{Yang,L}. It can be analogically conjectured
that one of the two ground states for the case of N odd has
$S_z=1/2$.
Note that in the interval $u \in (\pi/3,\>2\pi/3) $,
   the weights of the 6-vertex model (\ref{weights})
   are positive.  
   Consequently the components of the ground state vector are
   non-negative due to the Perron-Frobenius theorem.
   
  In collaboration with
 A. V. Razumov~\cite{RS}, we have found explicit values for
 these components up to $N=17$. They are all positive.
 Hence, if one of the two ground state vectors has $S_z=1/2$
 then we have identified it. Further we conjecture that 
 the  solution $Q(u)$ we found corresponds to the ground state
 for $N > 17$ as well.
 I believe  that using ODE (\ref{ODE})
 for $f(u)$ it is possible to describe the distribution 
 of the Bethe parameters and to prove this conjecture.
 
 For even N, the corresponding solution  $Q(u)$ has   
 degree  $N/2+1$ and there is no Bethe vector that 
ensure the simple eigenvalue we have discussed.

Now we return to the main calculations.
Firstly, we only know the largest eigenvalue
of the transfer matrix $T(u)=\sin^N u$ for $\eta=2\pi/3$.
It is remarkable that the knowledge of the second independent
solution $P(u)$ allows us to find the derivative of this 
eigenvalue w.r.t. $\eta$ and thus the simplest correlations.

Now we consider Baxter's T-Q equation (\ref{TQ}).
It can be interpreted  as a discrete version of
 a second order differential equation, so we can express its
 coefficients
in terms of two independent solutions~\cite{KLWZ,PRST}:
\begin{eqnarray}
&\sin ^N u=P(u+\eta /2) Q(u-\eta /2)
-P(u-\eta /2) Q(u+\eta /2), \nonumber \\
&T(u)=P(u+\eta) Q(u-\eta)-P (u-\eta) Q(u+\eta ). \nonumber
\end{eqnarray}
Using these relations similarly to paper~\cite{FSZ2}, we can find
the T-matrix derivative w.r.t $\eta$ 
\begin{eqnarray}
\label{tder}
&T_{\eta}^{\prime}(u)|_{\eta=2\pi/3}=
 \frac{3}{2}\biggl\{
 P(u+\pi/3)Q^{\prime}(u-\pi/3)
 - P^{\prime}(u+\pi/3) Q(u-\pi/3)+ \nonumber \\
&+ P(u-\pi/3)Q^{\prime}(u+\pi/3)
 - P^{\prime}(u-\pi/3) Q(u+\pi/3)\biggr\}. 
\end{eqnarray}

The derivative of the last equation w.r.t. the spectral parameter
u is 
\begin{eqnarray}
\label{t2der}
&T_{\eta u}^{\prime \prime}(u)|_{\eta=2\pi/3}=
 \frac{3}{2}\biggl\{
 P(u+\pi/3)Q^{\prime \prime}(u-\pi/3)
 - P^{\prime \prime}(u+\pi/3) Q(u-\pi/3)+ \nonumber \\
&+ P(u-\pi/3)Q^{\prime \prime}(u+\pi/3)
 - P^{\prime \prime}(u-\pi/3) Q(u+\pi/3)\biggr\}. 
\end{eqnarray}

 Let us use these derivatives.
 Averaging (\ref{HT}) over the ground state
we obtain the energy per site
that relates the correlations to the logarithmic
derivative of the largest eigenvalue of the
transfer matrix  w.r.t spectral parameter u 
\begin{equation}
   \label{ET}
    E_0(\eta)/N=-2 C^{\perp}_1
  -\cos \eta \> C^{\parallel}_1 = 
  \cos \eta - 
  \frac{2 \sin \eta}{N} \biggl(\frac
  {T^{\prime}_u}{T} \biggr)_{u=\eta/2}.
\end{equation}
For  $\eta=2\pi/3$  
and $T(u)=\sin ^N u$  this last equation is reduced to
\begin{equation}
\label{ququ}
E_0/N=-2 C^{\perp}_1(\Delta=-1/2)
  +\frac{1}{2}\>C^{\parallel}_1(\Delta=-1/2) = -\frac{3}{2}. 
\end{equation}
It is difficult to find two unknowns via one equation.
However, let us differentiate equation (\ref{ET}) w.r.t $\eta$.
$H$ is a Hermitian  operator and so one may 
ignore the $\eta$ dependence 
of the eigenvector. Hence, differentiating equation (\ref{ET}),
the matrix element of the Hamiltonian,
we can ignore $\eta$ dependence of the correlations
$$
  E^{\prime}_0(\eta)= 
  N \sin \eta \> C^{\parallel}_1 = 
   -N \sin \eta - 2 \biggl\{\cos \eta \biggl(\frac
  {T^{\prime}_u}{T} \biggr)+ 
  \sin \eta \frac{d}{d\eta}\biggl(\frac
  {T^{\prime}_u}{T} \biggr)\biggr\}_{u=\eta/2}.
$$
Jimbo and Miwa~\cite{JM} used this method for calculation
of the correlators $C^{\parallel}_1$
and $C^{\perp}_1$ in the thermodynamic limit.

Replacing $\eta=2\pi/3$ we  obtain 
the last but one formula for the correlation:  
$$
 C^{\parallel}_1(\Delta=-1/2) = 
  -\frac{1}{3}-\frac{2}{N}\frac{d}{d \eta}\biggl\{\biggl(\frac
  {T^{\prime}_u}{T} \biggr)_{u=\frac{\eta}{2}}
  \biggr\}_{\eta=\frac{2\pi}{3}}
  =1+\biggl(\frac{2}{\sqrt 3}\biggr)^{N+1}\biggl(T^{\prime}_{\eta}
  -\frac{\sqrt 3}{N}T^{\prime \prime}_{\eta u}\biggr)_
  {u=\frac{\pi}{3},\eta=\frac{2\pi}{3}}.
$$
The problem is solved in principle.
Due to formulas (\ref{tder}) and (\ref{t2der}),
we can express the correlations in terms of the two independent
solutions $Q(u)$ and $P(u)$ which are given (up to the 
factor $\sin^N u$ ) by the formulas (\ref{f}) and (\ref{g}).  
\bf The final answer \rm is
$$
<\sigma^z_j \sigma^z_{j+1}> \equiv C^{\parallel}_1(\Delta=-1/2) = 
- \frac{1}{2}+\frac{3}{2(2n+1)^2} =-\frac{1}{2}+\frac{3}{2N^2}.
$$
Detailed calculations will be published elsewhere.

Taking into account that the action of the operators
$\sigma_j^z\sigma_{j+1}^z$
depends upon the relative orientation of the neighbouring spins,
we can easily convert the last formula 
into a formula for the average number of "strings", i.e. 
clusters of spins with the same orientation:
 $$<N_{string}> = \frac{3}{4}(N-\frac{1}{N}).$$  
Using equation (\ref{ququ}),
we get the second correlation:
$$C^{\perp}_1(\Delta=-1/2)=
<\sigma_j^x\sigma_{j+1}^x>=<\sigma_j^y\sigma_{j+1}^y>=
\frac{5}{8}+\frac{3}{8N^2}.
$$

The obtained results can be generalized in different ways.
Let us discuss some possibilities.

First of all let us note that all
matrix elements of the Hamiltonian (\ref{H1})
are integers or halfintegers.
The ground state energy is a halfinteger as well.
It is not surprising that the normalization of the 
eigenvector can be chosen so that all its components 
are integers. This helps us to make calculations and using
Mathematica we have  explicitly found eigenvectors for $N \le 17$.
The obtained information is contained in 
publication~\cite{RS}.   
Let us mention only one result, related to the correlations,
which are called Probabilities of Formation of Ferromagnetic
String
\cite{KIB}.
All data are in agreement with the conjectured formula:
$$
\frac{<a_1\>a_2\>\dots\>a_{k-1}>}{<a_1\>a_2\>\dots\>a_{k}>}
=\frac{(2k-2)!\>(2k-1)!\>(2n+k)!\>(n-k)!}{(k-1)!\>(3k-2)!\>(2n-k+1)!\>
(n+k-1)!},
$$
where $a_j=(1+\sigma^z_{j})/2$

Many remarkable connections between 
the wave function components are noticeable.
As it turns out, the ratio of the largest component 
to the smallest one is equal to the number of AMS
(\it alternating sing matrices). 
\rm
The wonderful history of these numbers 
already has interweaved with the 6-vertex
model (see, for example, \cite{BP}).
Probably this is just the top of an iceberg.

Second one can try to consider the elliptic case.

Thirdly we can consider the inhomogeneous 6-vertex model.

I would like to mention that the necessity to distinguish between even 
and odd chains was remarked upon by Faddeev and Takhtajan~\cite{FT}
for the XXX chain, Baake et al. for XXZ~\cite{BCR},
Bugrij~\cite{Bu} for the Ising model and so on. 
In papers~\cite{DMN} authors derived benefit from consideration 
of XXZ spin chains with an odd number of sites.
As far as more recent results are concerned I want to mention our
paper~\cite{PRST}, where it was noticed that properties of 
Bethe equations depend on the parity of spin chain length. 
Papers of Schnack et al.~\cite{Sch} contain numeric investigation 
for the  spin 1/2,1,...,5/2 XXX chain and demonstrate the peculiarities 
of odd N.
Finally I would like to mention very recent Albertini paper~\cite{Al}
where author considers spin chains with open boundaries
and stresses that antiferromagnetic quantum spin chains
can in principle have very different properties according to the
parity of their length. 

\section*{Acknowledgments}
I would like to thank M. Batchelor, R. Baxter, V. Bazhanov,
A. Belavin, A. Bugrij,
 B. Feigin, T. Miwa, G. Pronko, V. Pugai, 
A. Razumov, S. Sergeev, G. Sierra and N. Slavnov
for useful discussions.
I also indebted to M. Henkel for bringing 
reference \cite{HS} to my attention. In that paper the authors
mention the special
features of the XXZ spin chain for $\Delta=-1/2$.
At last it is a pleasure to thank B. M. McCoy for sending me 
a list of references relating to
the derivative formula following (17)~\cite{RF,Mc}. 

This work is supported in part by RBRF--98--01--00070
and INTAS--96--690.


\begin{thebibliography}{XX}
\bibitem{BXYZ}
R. J. Baxter,
{\it Ann.\ Phys.}~70 (1972), 323
\bibitem{Baxter}
R. J. Baxter,
{\it Adv.\ Stud.\ in \ Pure\ Math.}~19 (1989), 95
\bibitem{ABBBQ}
F. C. Alcaraz, M. N. Barber, M. T. Batchelor, R. J. Baxter and 
 G. R. W. Quispel,
{\it J.\ Phys.}~A20 (1987), 6397
\bibitem{HMRS}
H. Hinrichsen, P. P. Martin, V. Rittenberg and  M. Scheunert,
{\it Nucl.\ Phys.}~B415 (1994), 533
\bibitem{MG}
Miguel A. Martin-Delgado and German Sierra
{\it Phys. \ Rev. \ Lett.}~76 (1996), 1146 
\bibitem{FSZ}
V. Fridkin, Yu. G. Stroganov and  Don Zagier,
{\it J.\ Phys.}~A33 (2000), L121
\bibitem{PS}
J. H. H. Perk and C. L. Schultz
{\it Phys.\ Lett.}~A84 (1981), 407
\bibitem{FSZ2}
V. Fridkin, Yu. G. Stroganov and  Don Zagier,
{\it J.\ of \ Stat.\ Phys.}~102 (2001), February issue
\bibitem{QISM}
P. P. Kulish and E. K. Sklyanin, 
{\it Phys. \ Lett.}~A70 (1979), 461;
\bibitem{QISM2}
L. A. Takhtajan and L. D. Faddeev,
{\it UMN}~34 ü5 (1979), 13
\bibitem{Yang}
C. N. Yang and C. P. Yang,
{\it Phys. \ Rev.}~150 (1966), 321;
{\it Phys. \ Rev.}~150 (1966), 327
\bibitem{L}
E. H. Lieb,
{\it Phys. \ Rev.}~162 (1967), 162;
{\it Phys. \ Rev. \ Lett. }~18 (1967), 1046;
{\it Phys. \ Rev. \ Lett. }~19 (1967), 108
\bibitem{RS}
A. V. Razumov and Yu. G. Stroganov,
In preparation
\bibitem{KLWZ}
I. Krichiver, O. Lipan, P. Wiegmann and A. Zabrodin,
{\it Commun.\ Math.\ Phys.}~188 (1997), 267
\bibitem{PRST}
G. P. Pronko and  Yu. G. Stroganov,
{\it J.\ Phys.}~A32 (1999), 2333
\bibitem{JM}
M. Jimbo, T. Miwa,
{\it J. \ Rhys. }~A29 (1996), 2923
\bibitem{KIB}
V. E. Korepin, A. G. Izergin and N. M. Bogoliubov,
{\it Quantum Inverse Scattering Method, Correlation Functions
and  Algebraic Bethe Ansatz,}~Cambridge University Press, 1993
\bibitem{BP}
D. Bressoud and J. Propp,
{\it Notices \ of the \ AMS}~46 (1999), 637
\bibitem{FT}
L. D. Faddeev and L. A. Takhtajan,
{\it Phys. \ Lett.}~A85 (1981), 375
\bibitem{BCR}
M. Baake, P. Christe and V. Rittenberg,
{\it Nucl. \ Phys.}~B300 (1988), 637
\bibitem{Bu}
A. Bugrij,
Private communication
\bibitem{HS}
M. Henkel and U. Schollwock,
Universal finite-size scaling amplitudes in anisotropic scaling,
cond-mat/0010061
\bibitem{BS}
A. A. Belavin and Yu. G. Stroganov,
{\it Phys.\ Lett.}~B466 (1999), 281
\bibitem{Su}
B. Sutherland,
{\it Phys. \ Rev. \ Lett. }~19 (1967), 103
\bibitem{ABB}
F. C. Alcaraz, M. N. Barber and M. T. Batchelor,
{\it Ann.\ Phys. (NY) }~182 (1988), 280
\bibitem{Sch}
K. B\"arwinkel, H.-J. Schmidt and J. Schnack,
{\it J.\ Magn. \ Magn. \ Mater. }~220 (2000), 227;
J. Schnack,
{\it Phys. \ Rev. }~B62 (2000), December issue
\bibitem{DMN}
A. Doikou, L. Mezincescu and R. I. Nepomechie,
{\it J. \ Phys. }~A30 (1997), L507; 
{\it Mod. \ Phys. \ Lett. }~A12 (1997), 2591 
\bibitem{Al}
G. Albertini,
Is the purely biquadratic spin 1 chain always massive?,
cond-mat/0012439
\bibitem{RF}
R.P. Feynman, 
{\it Phys. \ Rev. }~56 (1939), 340
\bibitem{Mc}
J.D. Johnson, S. Krinsky and B.M. McCoy,
{\it Phys. \ Rev. }~A8 (1973), 2526;
M. Gaudin. B.M. McCoy and T.T. Wu, 
{\it Phys. \ Rev. }~23D (1981), 417
\end{thebibliography}
\end{document}